\documentclass[11pt]{article}
\usepackage{amssymb}
\usepackage{graphics}
\usepackage{epsfig}
\usepackage{a4wide}
\let\jnfont=\rm
\def\NPB#1,{{\jnfont  Nucl.\ Phys.\ B }{\bf #1},}
\def\PLB#1,{{\jnfont Phys.\ Lett.\ B }{\bf #1},}
\def\EPJC#1,{{\jnfont Euro.\ Phys.\ J.\ C }{\bf #1},}
\def\PRD#1,{{\jnfont  Phys.\ Rev.\ D }{\bf #1},}
\def\PRL#1,{{\jnfont Phys.\ Rev.\ Lett.\ }{\bf #1},}
\def\MPLA#1,{{\jnfont Mod.\ Phys.\ Lett.\ A }{\bf #1},}
\def\JPG#1,{{\jnfont J.\ Phys.\ G}{\bf #1},}
\def\CTP#1,{{\jnfont Commun.\ Theor.\ Phys.\ }{\bf #1},}
\def\JHEP#1,{{\jnfont  JHEP }{\bf #1},}

\def\p_slash{\not{\hbox{\kern-2.1pt $p$}}}
\def\k_slash{\not{\hbox{\kern-2.1pt $k$}}}
\def\E_slash{\not{\hbox{\kern-2.1pt $E$}}}
\begin{document}

\title{ Flavor-Changing Top Quark Rare Decays in the Littlest Higgs Model with T-Parity } \vspace{3mm}

\author{{Hou Hong-Sheng}\\
{\small Ottawa-Carleton Institute for Physics, Department of Physics, Carleton University,}\\
{\small       Ottawa, K1S 5B6, Canada} }

\date{}
\maketitle \vskip 12mm

\begin{abstract}
We analyze the rare and flavor changing decay of the top quark
into a charm quark and a gauge boson in the littlest Higgs model
with T-parity (LHT). We calculate the one-loop level contributions
from the T-parity odd mirror quarks and gauge bosons. We find that
the decay $t\rightarrow c V,~(V=g,\gamma,Z)$ in the LHT model can
be significantly enhanced relative to those in the Standard Model.
Our numerical results show that the top quark FCNC decay can be as
large as $Br(t\rightarrow cg)\sim 10^{-2}$, $Br(t\rightarrow
cZ)\sim 10^{-5}$ and $Br(t\rightarrow c\gamma)\sim 10^{-7}$ in the
favorite parameter space in the LHT model.
\par
\end{abstract}

\vskip 5cm

{\large\bf PACS: 14.65.Ha,12.60.-i,12.15.Mm}

\vfill \eject

\baselineskip=0.32in

\renewcommand{\theequation}{\arabic{section}.\arabic{equation}}
\renewcommand{\thesection}{\Roman{section}.}
\newcommand{\nb}{\nonumber}
\newcommand{\be}{\begin{equation}}
\newcommand{\ee}{\end{equation}}

\makeatletter      % '@' is now a normal "letter" for TeX
\@addtoreset{equation}{section}
\makeatother       % '@' is restored as a "non-letter" character for TeX

\par
\section{Introduction}
\par
The detailed study of the dynamics of top quark production and
decay is an important objective of experiments at the Tevatron,
CERN Large Hadron Collider (LHC) and a possible International
Linear $e^+e^-$ Collider (ILC). The enormous production rates for
the top quark at the future colliders, reaching $10^7$ $t\bar{t}$
pairs per year at the LHC with luminosity of $10 fb^{-1}/y$ and
$10^5$ per year at a $500$ GeV $e^+e^-$ Collider with luminosity
of $100 fb^{-1}/y$, will allow to perform precision studies of top
quark properties. Due to the large mass of the top quark, it may
be more sensitive to new physics than other fermions and it may
serve as a window to probe new physics beyond the Standard Model
(SM).

In the SM, the flavor changing neutral current (FCNC) decays of
the top quark, $t \rightarrow cV$ with $V=\gamma,Z,g$, are absent
at tree level and are extremely suppressed at one-loop level due
to the GIM mechanism. Their branching ratios predicted in the SM
are of $O(10^{-10})$ or smaller \cite{Gprd1473,Bplb401}, far below
the detectable level at the Colliders. The top quark FCNC decays
have been extensively studied in various extensions of the SM ,
e.g. two-Higgs Doublet model (2HDM), left-right models, SUSY,
top-color assisted technicolor models (TC2), 331 models and models
with extra singlet quarks\cite{F0605003}. It has been realized
that the top quark FCNC decay modes can be enhanced by several
orders of magnitude in some scenarios beyond the SM and might fall
in the reach of the future Colliders.

The Little Higgs mechanism \cite{LH-origin,LH-review} offers an
alternative approach to SUSY for weakly coupled electroweak
symmetry breaking (EWSB) without fine-tuning. The most compact
implementation of the Little Higgs mechanism is known as the
Littlest Higgs model \cite{N0206021,H0301040}. In this model,
there are new vector bosons, a heavy top quark and a triplet of
heavy scalars in addition to SM particles. The original Little
Higgs models suffer strong constraints from electroweak precision
data\cite{constraints}. To solve this problem, a ${\cal Z}_2$
discrete symmetry named "T-parity" (analogous to R-parity in SUSY)
is introduced in Refs. \cite{LHT,I0409025}. In the Littlest Higgs
model with T parity (LHT), a large part of the model parameter
space is consistent with data, and values of the symmetry breaking
scale $f$ can be as low as 500 GeV \cite{J0506042}.

The LHT model requires the introduction of "mirror fermions" for
each SM fermion doublet. The mirror fermions are odd under
T-parity and can be given large masses. We can introduce new
flavor interactions in the mirror quark sector that could have a
very different pattern from the ones present in the SM. In Refs.
\cite{J0512169,M0605214,M0610298}, the impact of the mirror
fermions on FCNC processes such as neutral meson mixing and rare
$K,B$ meson decays in the LHT model are studied in detail. In this
paper, we will concentrate on the top quark FCNC decays $t
\rightarrow cV$ in the LHT model.

The structure of this paper is as follows: In Sec. II, we briefly
review the ingredients of the LHT model that are relevant to our
calculation. In Sec. III, numerical analysis of $t \rightarrow cV$
in the LHT is presented. Finally, we give a short conclusion in
Sec. IV.
 \vskip
5mm
\section{The Model}
A detailed description of the LHT model can be found in the
literature \cite{J0411264,J0512169}, here we only review the
ingredients which are relevant to the analysis in this paper.

\subsection{ Scalar sector}
The Littlest Higgs model embeds the electroweak sector of the
standard model in an $SU(5)/SO(5)$ non-linear sigma model. It
begins with an $SU(5)$ global symmetry with a locally gauged
subgroup $[SU(2) \times U(1)]^2$. The $SU(5)$ symmetry is
spontaneously broken down to $SO(5)$ via a vacuum expectation
value (VEV) of order $f$. At the same time, the $[SU(2) \times
U(1)]^2$ gauge symmetry breaks to its diagonal subgroup $SU(2)_L
\times U(1)_Y$ which is identified as the SM electroweak gauge
group.

From the $SU(5)/SO(5)$ breaking, there arise 14 Goldstone bosons
which are described by the "pion" matrix $\Pi$, given explicitly
by
\be\label{Pi}\addtolength{\arraycolsep}{3pt}\renewcommand{\arraystretch}{1.3}
 \Pi=\left(\begin{array}{ccccc}
-\frac{\omega^0}{2}-\frac{\eta}{\sqrt{20}} &
-\frac{\omega^+}{\sqrt{2}} &
  -i\frac{\pi^+}{\sqrt{2}} & -i\phi^{++} & -i\frac{\phi^+}{\sqrt{2}}\\
-\frac{\omega^-}{\sqrt{2}} &
\frac{\omega^0}{2}-\frac{\eta}{\sqrt{20}} &
\frac{v+h+i\pi^0}{2} & -i\frac{\phi^+}{\sqrt{2}} & \frac{-i\phi^0+\phi^P}{\sqrt{2}}\\
i\frac{\pi^-}{\sqrt{2}} & \frac{v+h-i\pi^0}{2} &\sqrt{4/5}\eta &
-i\frac{\pi^+}{\sqrt{2}} & \frac{v+h+i\pi^0}{2}\\
i\phi^{--} & i\frac{\phi^-}{\sqrt{2}} & i\frac{\pi^-}{\sqrt{2}} &
-\frac{\omega^0}{2}-\frac{\eta}{\sqrt{20}} & -\frac{\omega^-}{\sqrt{2}}\\
i\frac{\phi^-}{\sqrt{2}} &  \frac{i\phi^0+\phi^P}{\sqrt{2}} &
\frac{v+h-i\pi^0}{2} & -\frac{\omega^+}{\sqrt{2}} &
\frac{\omega^0}{2}-\frac{\eta}{\sqrt{20}}
\end{array}\right).
\ee where it consists of a doublet $H$ and a triplet $\Phi$ under
the unbroken $SU(2)_L \times U(1)_Y$ group which are
given by \be H= \left(\begin{array}{c} -i\frac{\pi^+}{\sqrt{2}} \\
\frac{v+h+i\pi^0}{2} \end{array}\right),~~~~~~\Phi=
\left(\begin{array}{cc} -i\phi^{++} &
-i\frac{\phi^+}{\sqrt{2}}\\-i\frac{\phi^+}{\sqrt{2}} &
\frac{-i\phi^0+\phi^P}{\sqrt{2}}
\end{array}\right). \ee
Here, $H$ plays the role of the SM Higgs doublet, $h$ is the
physical Higgs field and $v\simeq246$ GeV. The fields $\eta$ and
$\omega$ are eaten by heavy gauge bosons when the $[SU(2) \times
U(1)]^2$ gauge group is broken down to $SU(2)_L\times U(1)_Y$,
whereas the $\pi$ fields are absorbed by the standard model $W/Z$
bosons after EWSB. The field $h$ and $\Phi$ remain in the
spectrum.

In the LHT model, a T-parity discrete symmetry is introduced to
make the model consistent with the data. Under the T-parity, the
fields $\Phi$, $\omega$ and $\eta$ are odd, and the SM Higgs
doublet $H$ is even. The Goldstones $\omega$ and $\eta$ are
present in our analysis.

\subsection{Gauge boson sector}
In the LHT model, the T-even gauge boson sector consists only of
the SM gauge bosons \be W_L^\pm, ~~ Z_L,~~A_L \ee with masses
given to lowest order in $v/f$ by \be
M_{W_L}=\frac{gv}{2},~~M_{Z_L}=\frac{M_{W_L}}{\cos \theta_W},~~
M_{A_L} =0, \ee where $\theta_W$ is the weak mixing angle, and $g$
is the SM $SU(2)$ gauge couplings.

The T-odd gauge boson sector consists of three heavy "partners" of
the SM gauge bosons \be W_H^\pm, ~~ Z_H,~~A_H \ee with masses
given to lowest order in $v/f$ by \be M_{W_H}=gf ,~~M_{Z_H}=gf,~~
M_{A_H} =\frac{g^\prime f}{\sqrt{5}}, \ee where $g^\prime$ is the
SM $U(1)$ gauge couplings. All these heavy gauge bosons will be
involved in our analysis.

\subsection{Fermion sector}
The T-even fermion sector consists of the SM quarks, leptons and
an additional heavy quark $T_+$.

The T-odd fermion sector consists of three generations of mirror
quarks and leptons and an additional heavy quark $T_-$. The $T_-$
will not be present in our analysis for the reason discussed in
appendix A of Ref. \cite{M0605214}. Only the mirror quarks are
involved in this paper. We denote them by \be
\left(\begin{array}{c} u_H^1 \\ d_H^1 \end{array} \right),
~~\left(\begin{array}{c} u_H^2
\\ d_H^2 \end{array} \right),~~\left(\begin{array}{c} u_H^3 \\
d_H^3 \end{array} \right).\ee To the first order of $v/f$, their
masses satisfy \be m_{u_H^1}=m_{d_H^1}\equiv
m_{H1},~~m_{u_H^2}=m_{d_H^2}\equiv
m_{H2},~~m_{u_H^3}=m_{d_H^3}\equiv m_{H3}.\ee

\subsection{T-odd flavor mixing}
In the LHT, the mirror fermions open up a new flavor structure in
the model. As discussed in Ref. \cite{J0512169,M0605214,M0610298},
there are four CKM-like unitary mixing matrices in the mirror
fermion sector: \be V_{H_u},~~V_{H_d},~~V_{H_l},~~V_{H_{\nu}}. \ee
$V_{H_u}$ and $V_{H_{d}}$ are for the mirror quarks which are
present in our analysis. These mirror mixing matrices are involved
in the flavor changing interactions between SM fermions and T-odd
mirror fermions which are mediated by the T-odd heavy gauge and
Goldstone bosons ( $W_H,Z_H,A_H$ and $\omega^\pm,\omega^0,\eta$).
$V_{H_u}$ and $V_{H_{d}}$ satisfy the relation \be V_{H_u}^\dagger
V_{H_{d}} = V_{\rm CKM}. \ee  Following the method in
\cite{M0609284,M0610298}, we parameterize the $V_{Hd}$ with three
angles $\theta_{12}^d,\theta_{23}^d,\theta_{13}^d$ and three
phases $\delta_{12}^d,\delta_{23}^d,\delta_{13}^d$ \be V_{Hd}=
\left(\begin{array}{ccc}
c_{12}^d c_{13}^d & s_{12}^d c_{13}^d e^{-i\delta^d_{12}}& s_{13}^d e^{-i\delta^d_{13}}\\
-s_{12}^d c_{23}^d e^{i\delta^d_{12}}-c_{12}^d s_{23}^ds_{13}^d
e^{i(\delta^d_{13}-\delta^d_{23})} & c_{12}^d c_{23}^d-s_{12}^d
s_{23}^ds_{13}^d e^{i(\delta^d_{13}-\delta^d_{12}-\delta^d_{23})}
&
s_{23}^dc_{13}^d e^{-i\delta^d_{23}}\\
s_{12}^d s_{23}^d e^{i(\delta^d_{12}+\delta^d_{23})}-c_{12}^d
c_{23}^ds_{13}^d e^{i\delta^d_{13}} & -c_{12}^d s_{23}^d
e^{i\delta^d_{23}}-s_{12}^d c_{23}^d s_{13}^d
e^{i(\delta^d_{13}-\delta^d_{12})} & c_{23}^d c_{13}^d
\end{array}\right).
\ee The matrix $V_{Hu}$ is then determined through $V_{Hu}=V_{Hd}
V_{\rm CKM}^\dagger$.  The Feynman rules for the flavor violating
interactions which are involved in our analysis can be found in
Appendix B of Ref. \cite{M0610298}.

\section{$t \rightarrow cV$ in the LHT model}
In the LHT model, the flavor changing rare top decay $t
\rightarrow cV~~(V=\gamma,Z,g)$ can be induced by the interactions
between SM quarks and T-odd mirror quarks mediated by heavy T-odd
gauge bosons and Goldstone bosons at one-loop level. The relevant
Feynman diagrams are shown in Fig.1.

\begin{figure}[h]
\centering
\scalebox{0.9}{\includegraphics*[580,285][125,580]{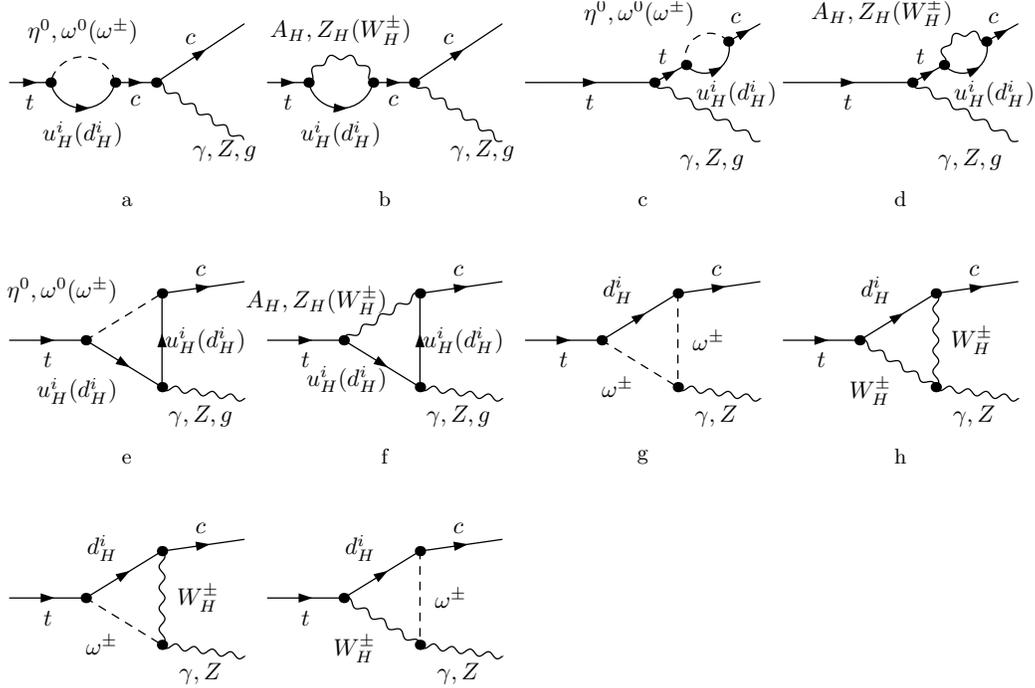}}
\caption{The feynman diagrams for $t \rightarrow c
V~(V=\gamma,Z,g)$ in the LHT.}
\end{figure}

The amplitude of the decay is given by \be {\cal M}(t\rightarrow
cV)=\bar{u}(p_2) V^\mu u(p_1) \epsilon_\mu(k,\lambda), \ee where
$p_1$, $p_2$ and $k$ are the momenta of the incoming top quark,
outgoing charm quark, and outgoing gauge boson, respectively.
$\epsilon_\mu(k,\lambda)$ is the polarization vector for the
outgoing gauge boson. At one-loop level, the diagrams in Fig.1
give rise to effective vertices $V^\mu$ which can written as
\begin{eqnarray}
V^\mu(tcZ)=i e [\gamma^\mu (P_L A_L^Z+P_R A_R^Z) + k_\mu
\sigma^{\mu \nu}(P_L B_L^Z +P_R B_R^Z) ] \nb \\
V^\mu(tc\gamma)=i e [\gamma^\mu (P_L A_L^\gamma+P_R A_R^\gamma) +
k_\mu
\sigma^{\mu \nu}(P_L B_L^\gamma +P_R B_R^\gamma) ] \nb \\
V^\mu(tcg)=i g_s T^a [\gamma^\mu (P_L A_L^g+P_R A_R^g) + k_\mu
\sigma^{\mu \nu}(P_L B_L^g +P_R B_R^g) ]
\end{eqnarray}
where $P_{R,L}=\frac{1}{2}(1\pm\gamma^5$, $\sigma^{\mu
\nu}=\frac{i}{2}[\gamma^\mu,\gamma^\nu]$ and $T^a$ are the
generators of $SU(3)_C$. The form factors $A_{L,R}, B_{L,R}$ are
the sums of the contributions from the diagrams in Fig.1. They
encode the loop functions and depend on the masses and momenta of
the external and internal quarks, heavy quarks, gauge bosons and
heavy gauge bosons. For simplicity, we omit the explicit
expressions for these form factors. The feynman diagrams and the
corresponding amplitudes are generated by using {\it
FeynArts3}\cite{FeynArts}. We have added the relevant Feynman
rules of the LHT model in {\it FeynArts} package. In the
calculations of the one-loop diagrams we adopt the definitions of
one-loop integral functions as in Ref.\cite{loopintegral}. The
loop integral functions are calculated by using the formulas in
Ref.\cite{abcd}.

The new parameters  in the LHT which are relevant to our analysis
are \be f,~
m_{H1},~m_{H2},~m_{H3},~\theta_{12}^d,~\theta_{23}^d,~\theta_{13}^d,
~\delta_{12}^d,~\delta_{23}^d,~\delta_{13}^d \ee It is convenient
to consider several representative scenarios for the structure of
the $V_{Hd}$. In Ref.\cite{J0512169}, the constraints on the mass
spectrum of the mirror fermions have been investigated from the
analysis of neutral meson mixing in the $K$, $B$ and $D$ systems.
They found that a TeV scale GIM suppression is necessary for a
generic choice of $V_{Hd}$. However, there are regions of
parameter space where there are only very loose constraints on the
mass spectrum of the mirror fermions. We will study the
$t\rightarrow c V$ in the regions of parameter space where the
mass spectrum is not severely constrained in Ref.\cite{J0512169}.
We concentrate our study in the following two scenarios for the
structure of the $V_{Hd}$
\begin{itemize}
\item {\bf Case I}\hspace{.08in} $V_{Hd} = {\mathbf 1}$, $V_{Hu} =
V_{{\rm CKM}}^\dagger$ \item {\bf Case II}\hspace{.08in}
$s_{23}^d=1/\sqrt{2},s_{12}^d=0,s_{13}^d=0,\delta_{12}^d=0,\delta_{23}^d=0,\delta_{13}^d=0$
%\item
%{\bf Case III} $V_{Hu}^\dagger = V_{Hd}$, $V_{{\rm CKM}} =
%V_{Hu}^{\dagger 2} = V_{Hd}^2$
\end{itemize}
In both the cases, the constraints on the mass spectrum of the
mirror fermions are very relaxed.

Before presenting numerical the results of $t\rightarrow c V$ in
the LHT, we would like to comment on the sensitivity of the future
experiments to FCNC top quark decays. At the LHC with a
$100~fb^{-1}$ integrated luminosity, the branching ratios can be
measured as small as \cite{ATLAS} \begin{eqnarray} Br(t\rightarrow
cg) \leq 7.4\times 10^{-3}, \nb \\ Br(t\rightarrow cZ) \leq
1.1\times 10^{-4}, \nb \\ Br(t\rightarrow c\gamma) \leq 1.0 \times
10^{-4}\end{eqnarray}

\begin{figure}[thb]
\centerline{\scalebox{0.6}{\includegraphics*[495,80][70,415]{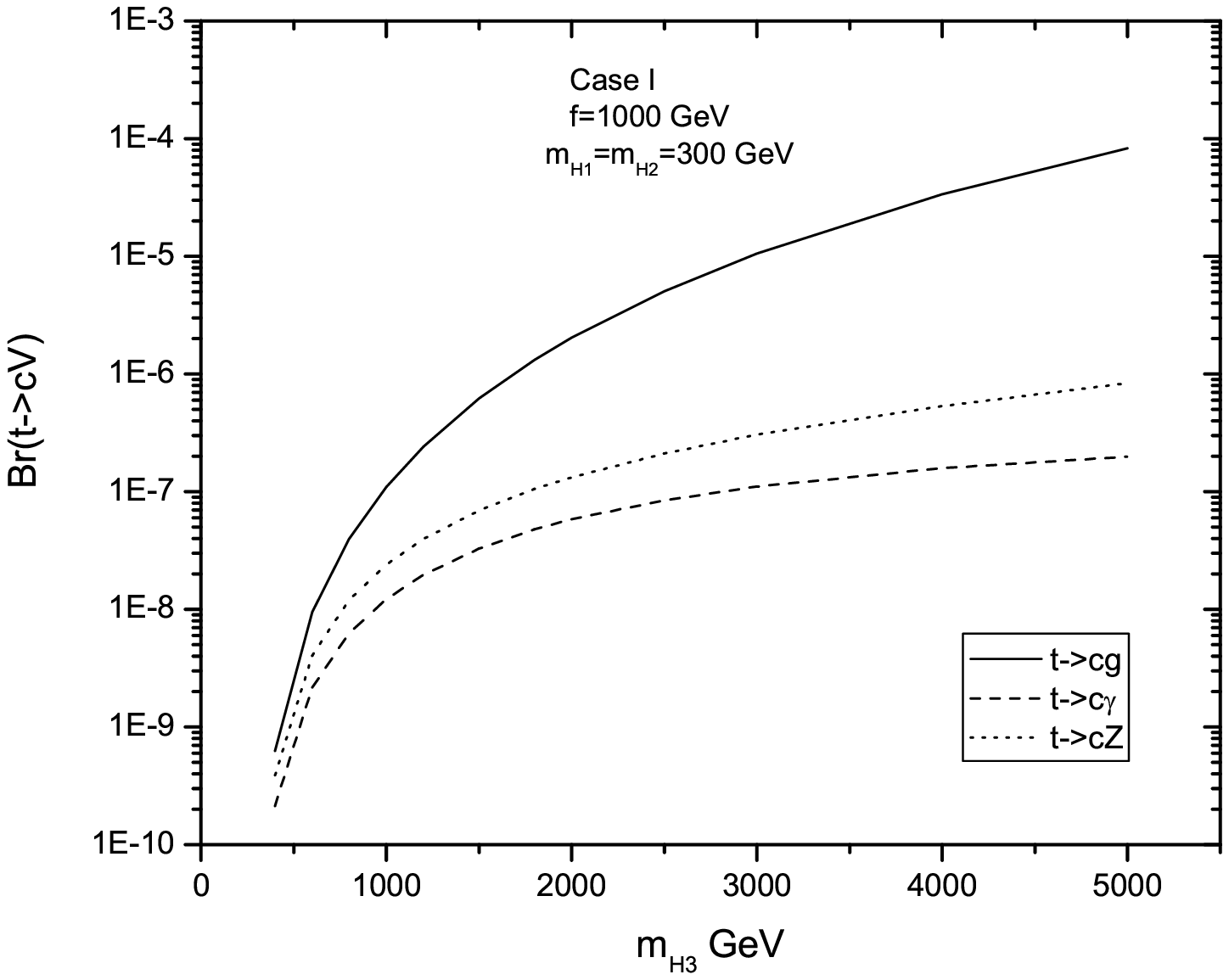}}\
\scalebox{0.6}{\includegraphics*[495,80][70,415]{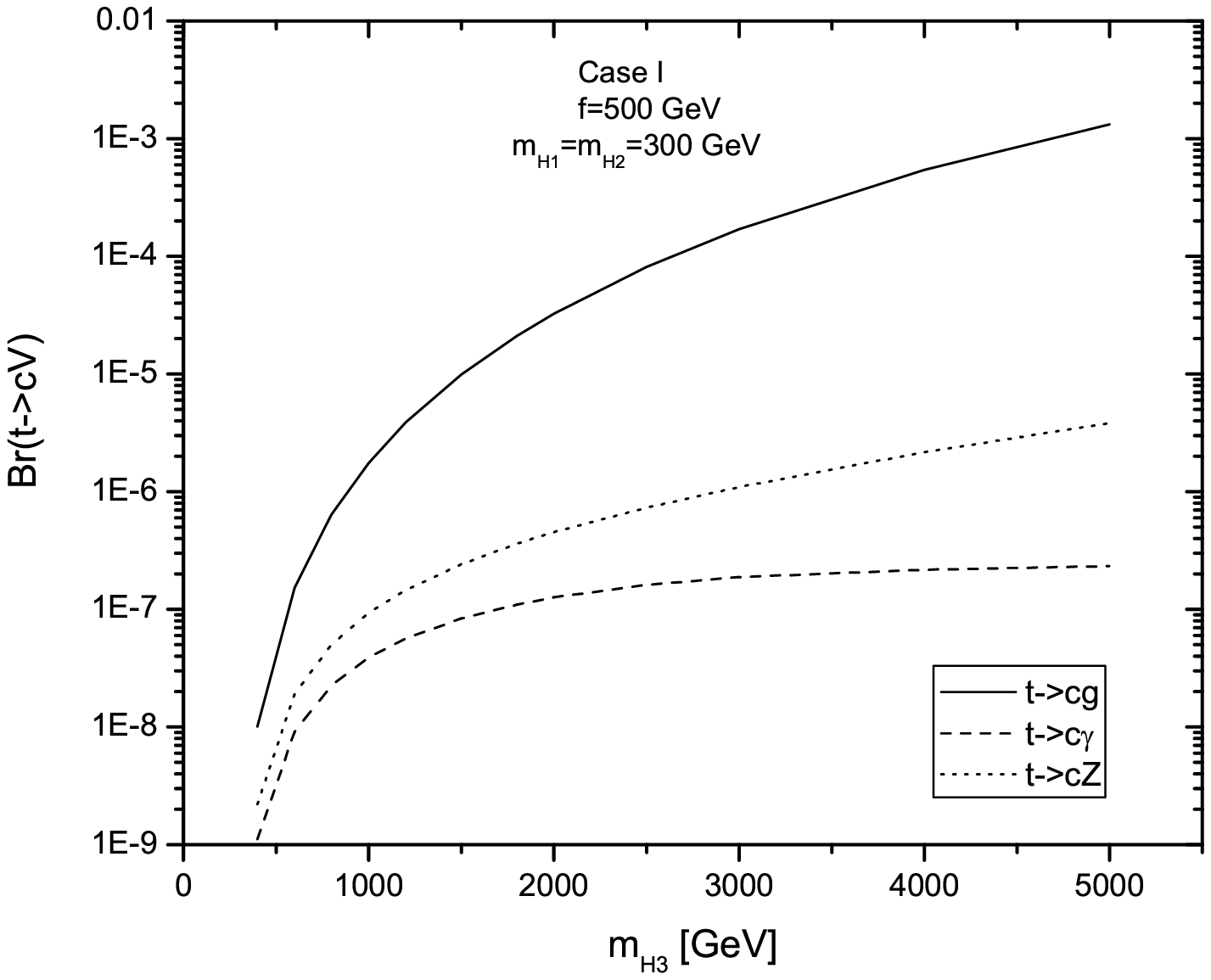}}}
\caption{{\bf Case I} $V_{Hd} = {\mathbf 1}$, $V_{Hu} = V_{{\rm
CKM}}^\dagger$. For the left figure, $f$ is taken to be 500 GeV.
For the right one, $f$ is equal to 1000 GeV.}
\end{figure}

Now we present the numerical results for $Br(t \rightarrow cV)$.
We take $m_t=171.4$ GeV, $m_Z = 91.187$ GeV, $m_W=80.4$ GeV,
$m_c=1.25$ GeV, $G_F=1.16639\times 10^{-5}{\rm (GeV)^{-2}}$,
$\alpha=1/128$ and $\alpha_s=0.107$\cite{pdg}. Because the total
width of
 the top-quark is dominated by the leading decay mode $t
 \rightarrow bW^+$, we define the branching ratio as
 \be
 Br(t \rightarrow cV)=\frac{\Gamma(t \rightarrow c V)}{\Gamma(t\rightarrow
 bW^+)}. \ee

In fig.2 we present the plot of the branching ratio of
$t\rightarrow c V~(V=g,\gamma,Z)$ as a function of the mass of the
third generation mirror quark for Case I. In this case we have
taken $V_{Hd} = {\mathbf 1}$, $V_{Hu} = V_{{\rm CKM}}^\dagger$,
the mixing in the down type gauge and Goldstone boson interactions
are absent. In this case there are no constraints on the masses of
the mirror quarks at one loop level from the $K$ and $B$ systems
and the constraints come only from the $D$ system. The constraints
on the mass of the third generation mirror quark are very weak
according to the analysis in Ref. \cite{J0512169}. From fig.2, we
can see that the branching ratios of $t\rightarrow c V$ rise very
fast with the increasing mass of the third generation mirror
quark. For $f=500$ GeV and $m_{H3}=5000$ GeV, the branching ratios
can reach
\begin{eqnarray}
Br(t\rightarrow c g)\approx 1.33\times 10^{-3} \nb \\
Br(t\rightarrow c Z)\approx 3.83\times 10^{-6}  \nb \\
Br(t\rightarrow c \gamma)\approx 2.33\times 10^{-7}.
\end{eqnarray}

\begin{figure}[thb]
\centerline{\scalebox{0.6}{\includegraphics*[450,20][20,370]{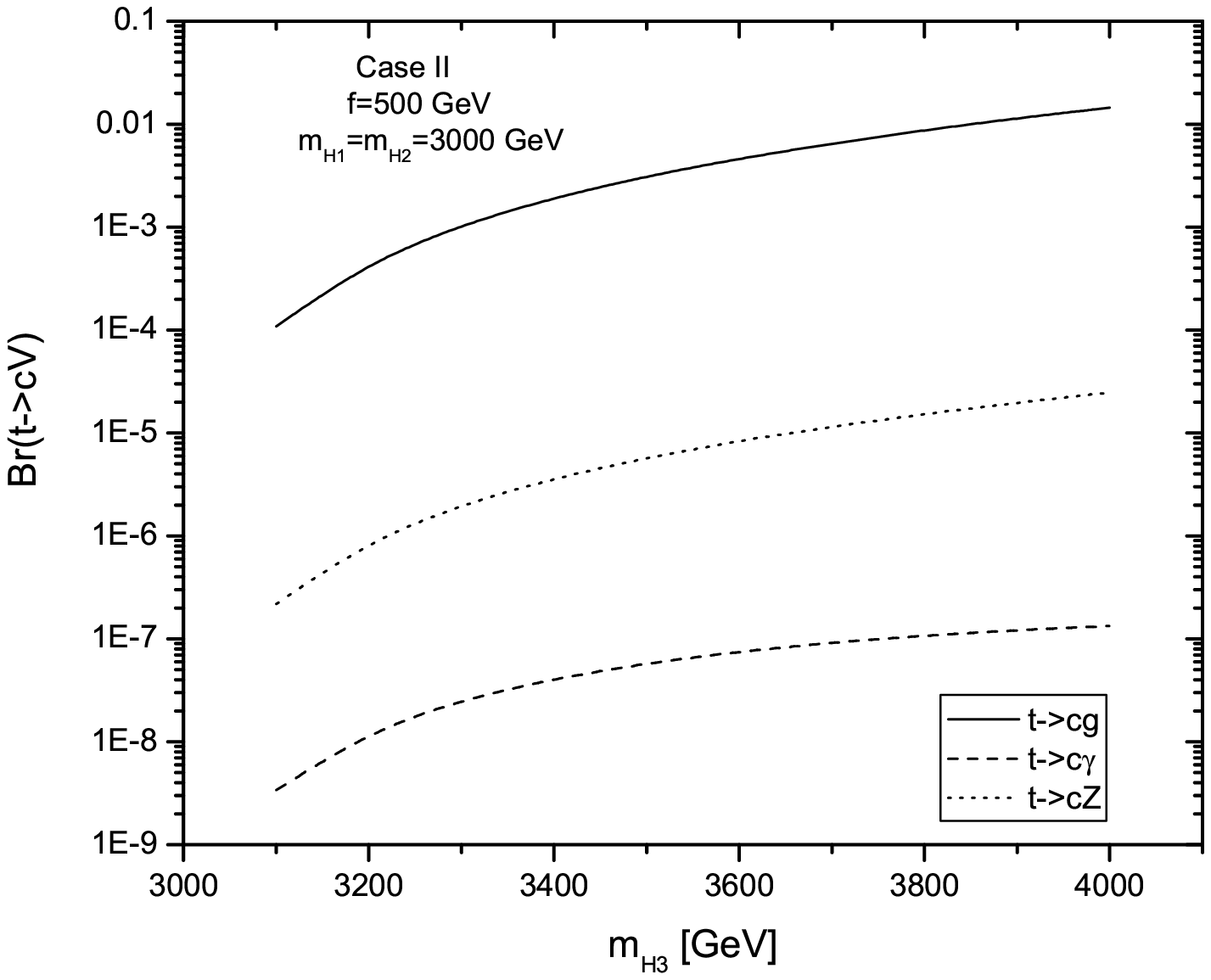}}\
\scalebox{0.6}{\includegraphics*[450,20][20,370]{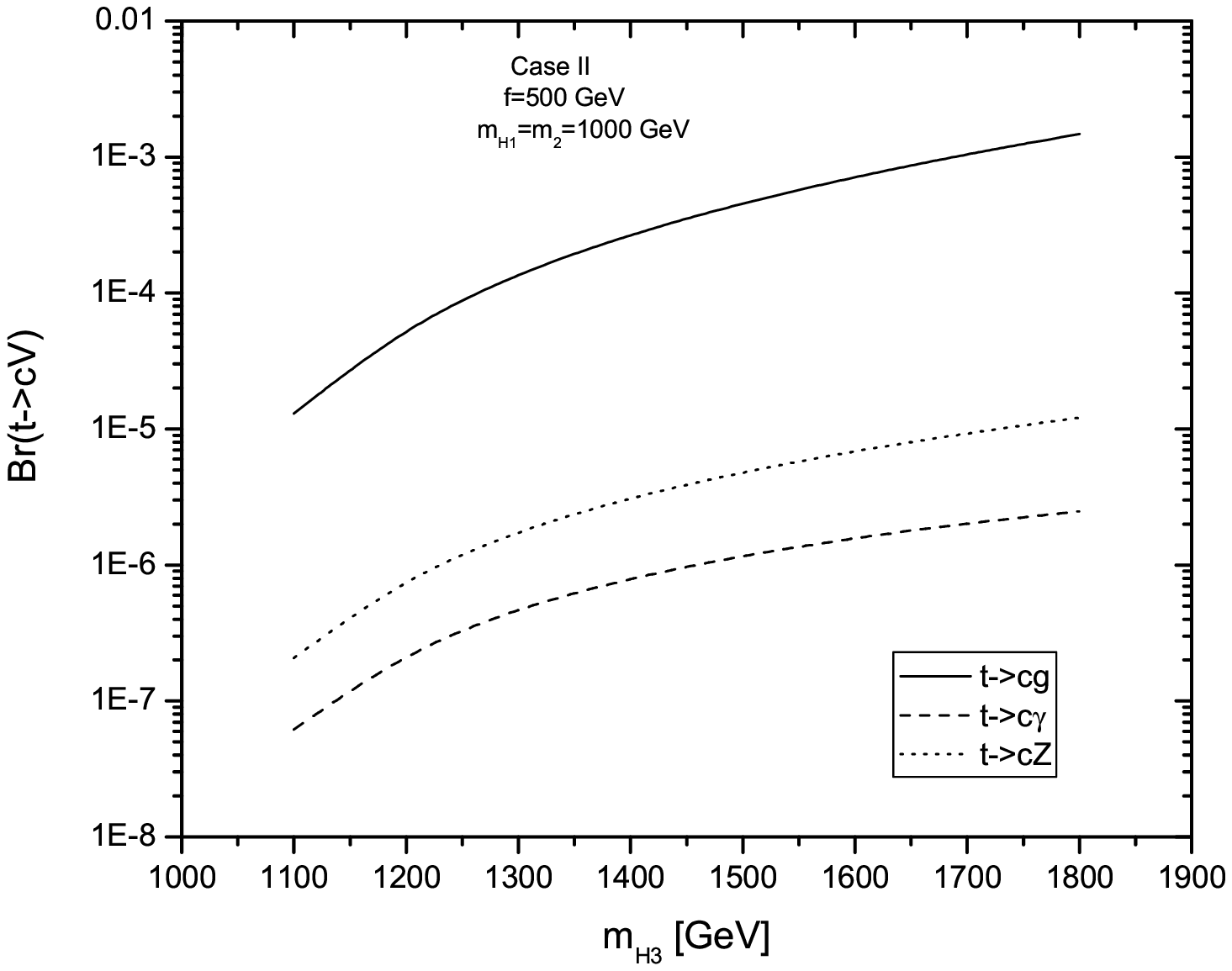}}}
\caption{{\bf Case II} $f=500$ GeV, $s_{23}^d=1/\sqrt{2}$, while
other angles are set to zero. For the left figure, $m_{H1}=m_{H2}$
is taken to be 1000 GeV. For the right one, $m_{H1}=m_{H2}$ is
equal to 3000 GeV.}
\end{figure}

In fig.3 the dependence of branching ratio of $t\rightarrow c V$
on $m_{H3}$ is presented for Case II. In this case, the
constraints from the $K$ and $B$ systems are also very weak.
Compared to Case I, the mixing between the second and the third
generations are enhanced with the choice of a bigger mixing angle
$s_{23}^d$. The branching ratios of $t\rightarrow c V$ are
significantly enhanced even with stricter constraints on the
masses of the mirror quarks from the $D$ system. For $f=500$ GeV,
$m_{H1}=m_{H1}=3000$ and $m_{H3}=4000$ GeV, the branching ratios
can reach
\begin{eqnarray}
Br(t\rightarrow c g)\approx 1.44\times 10^{-2} \nb \\
Br(t\rightarrow c Z)\approx 2.56\times 10^{-5}  \nb \\
Br(t\rightarrow c \gamma)\approx 1.33\times 10^{-7}.
\end{eqnarray}
In this case, the branching ratio of $t \rightarrow cg$ can reach
the detectable level of the LHC. For $t \rightarrow cZ$, the
branching ratio is slightly below the expected reach of the LHC.

 \vskip 5mm
\section{Conclusion}
We have calculated the one-loop contributions from the T-odd
quarks and gauge bosons to the rare and flavor changing decay of
the top quark in the LHT model. We find the branching ratios of
the $t \rightarrow c V, ~(V=g,\gamma,Z)$ in the LHT model can be
significantly enhanced relative to those in the SM. In our
numerical analysis, we have chosen two favorite scenarios for the
structure of the mixing matrix $V_{Hd}$ in the mirror fermion
sector. For both of these cases, the constraints on the mass
spectrum of the mirror quarks from the data of neutral meson
mixing are very weak. Our numerical results show that the
branching ratios of top quark decay into a charm quark and a gauge
boson can reach to $Br(t\rightarrow cg)\sim 10^{-2}$,
$Br(t\rightarrow cZ)\sim 10^{-5}$ and $Br(t\rightarrow
c\gamma)\sim 10^{-7}$ for optimistic parameter choice in the LHT
model.

\vskip 5mm

\section*{Acknowledgments} I would like to thank P.
Kalyniak, H. Logan and S. Godfrey for helpful discussions and
suggestions. I thank H. Logan and S. Godfrey for reading the
manuscript. This research was funded by the Natural Sciences and
Engineering Research Council of Canada.

\end{document}